\documentclass[final,5p,times,twocolumn,sort&compress]{elsarticle}%%%twocolumn
\usepackage{epsfig}
\usepackage{amsmath}
\usepackage{url}
\usepackage[colorlinks,linkcolor=blue,anchorcolor=blue,citecolor=blue,urlcolor=blue]{hyperref}
\usepackage{graphicx}
\usepackage{lineno}
\usepackage{color}
\usepackage{ulem}
\usepackage{float}
\usepackage{stfloats}
\usepackage[marginal]{footmisc}

\journal{Physics Letter B}
\begin{document}

\begin{frontmatter}

\title{Constraining equation of state of nuclear matter by charge-changing cross section measurements of mirror nuclei}

\author[bh]{Jun-Yao Xu \fnref{1}}

\author[Lzu]{Zheng-Zheng Li\fnref{1}}

\author[bh]{Bao-Hua Sun\corref{cor1}}
\ead{bhsun@buaa.edu.cn}

\author[Lzu]{Yi-Fei Niu\corref{cor1}}
\ead{niuyf@lzu.edu.cn}

\author[INFN]{Xavier Roca-Maza\corref{cor1}}
\ead{xavier.roca.maza@mi.infn.it}

\author[RNC]{Hiroyuki Sagawa}
\author[bh]{Isao Tanihata}

%\date{\today}
%\cortext[cor0]{These authors contributed equally to this work.}
\cortext[cor1]{Corresponding Author}
\address[bh]{School of Physics, Beihang University, Beijing 100191, China}

\address[Lzu]{School of Nuclear Science and Technology, Lanzhou University, Lanzhou 730000, China}

\address[INFN]{Dipartimento di Fisica, Universit\`{a} degli Studi di Milano and INFN, Sezione di Milano, 20133 Milano, Italy}

\address[RNC]{RIKEN Nishina Center, Hirosawa 2-1, Wako, Saitama 351-0198, Japan}
\fntext[1]{These authors contributed equally to this work.}

\begin{abstract}
The nuclear symmetry energy plays a key role in determining the equation of state (EoS) of dense, neutron-rich matter, which connects the atomic nuclei with the hot and dense matter in universe, thus has been the subject of intense investigations in laboratory experiments, astronomy observations and theories. 
Various probes have been proposed to constrain the symmetry energy and its density dependence.
Currently, the extensive data yield already a good and consistent constraint to the symmetry energy ($E_\text{sym}(\rho)$) at saturation density, but do not yet give a consistent result of one critical EoS parameter, $L$, the density dependence of the symmetry energy.  
In this work, we report a new probe of $L$ at saturation density. 
A good linear correlation is found between $L$ and the charge changing cross section difference ($\Delta\sigma_\text{cc}$) of mirror nuclei $^{30}$Si-$^{30}$S for both the Skyrme-Hartree-Fock theory (SHF) and covariant (relativistic) density functionals (CDF). We found that the pairing effect for this mirror pair is essential to get a consistent correlation between $L$ and $\Delta\sigma_\text{cc}$ in both the SHF and CDF. 
Here, the cross sections are calculated on the same target and at the same energy using the zero-range optical-limit Glauber model.  
The linearity is found to be in the same precision as those found between $L$ and neutron skin thickness or proton radius difference.

\end{abstract}

%\pacs{23.50.+z,23.35.+g,27.70.+q,23.20.Lv,21.10.Tg}
%\maketitle

\begin{keyword}
Symmetry energy \sep Equation of state of nuclear matter \sep Charge-changing cross section \sep Mirror nuclei 

\end{keyword}
\end{frontmatter}

%%%\bibliographystyle{elsarticle-num}

%%% Start the text now
%%\linenumbers
%%\setrunninglinenumbers
%%\leftlinenumbers*

\section{Introduction}
\label{Section:Intro}

The symmetry energy of nuclear matter, $E_\text{sym}(\rho)$, is the energy required to create an isospin asymmetry between neutrons and protons. It plays a key role in understanding the equation of state (EoS) of dense, neutron-rich matter, and has 
a crucial impact in nucleus-nucleus collisions, in the structure and stability of atomic nuclei, and 
also in supernova explosions and neutron star stability~\cite{Stein2006Phys.Rep,LiBao2008Phys.Rep,Latti2007Phys.Rep,ShenH2020ApJ,XRoca2018PPNP,Burgi2021PPNP,Baldo2016PPNP,Latti2016PR}. 
The investigation of $E_\text{sym}(\rho)$ 
can be accomplished  from normal to light density region by compressing and heating up nuclear matter in high-energy heavy-ion collisions~\cite{Zhang2020FoP,LiHan2020PRL,XuHao2021PLB}, 
by  properties of finite nuclei~\cite{WanNi2016PRC,XRoca2018PPNP}, 
and by detecting the binary neutron-star merger event~\cite{Zhou2019ApJ,Baiot2019PPNP}.

The symmetry energy at saturation density, $\rho_0$, has been constrained to be around 30 MeV~\cite{Willi1966Nucl.Phys,Pomor2003PRC}. However, the density  dependence of symmetry energy, $L=3\rho{[\partial E_{sym}(\rho)}/{\partial \rho}]|_{\rho=\rho_0} $, remains to have a significant uncertainty. $L$ provides the dominant baryonic contribution to the pressure in neutron stars~\cite{Horow2001PRL} and affects the neutron skin thicknesses in heavy nuclei~\cite{Tsang2012PRC,Horow2014JPG}. Constraints of $L$ have been put on, e.g., using the neutron-skin thickness of finite nuclei~\cite{Typel2001PRC,Reinh2016PRC}, the dipole polarizability of finite nuclei ~\cite{Tamii2011PRL,Latti2013ApJ}, the radius of neutron stars ~\cite{stein2012PRL}, proton-elastic scattering~\cite{iida2003PLB}, charge-exchange reactions~\cite{Krasznahorkay1999,Krasznahorkay2013,Cao2015,Roca-Maza2016}, the nucleon global optical potentials~\cite{Chang2010PRC}, and also from heavy-ion collisions~\cite{WangY2020PLB} and astrophysical heavy pulsar masses~\cite{Essic2021PRL}. 
%\textcolor{red}{Meanwhile, it's completely determined by , which can be extracted directly from nucleon-nucleus scatterings, ($p,n$) charge-exchange reactions, a .}
The latest compilations give $L$ in the range of 40.5-61.9~\cite{Latti2013ApJ},
42.4-75.4~\cite{LiBao2008PLB} and 30.6-86.8 MeV~\cite{Oerte2017RevModPhys}. 
 
The neutron skin thickness, particularly for $^{208}$Pb, is of great contemporary interest 
due to the strong correlation between  $R_{\text{skin}}^{208}$ and 
the slope of the symmetry energy $L$~\cite{Brown2000PRL,Chenl2005PRC,Cente2009PRL,RocaM2011PRL}. 
The updated neutron skin thickness of $^{208}$Pb ($R_{\text{skin}}^{208}$) from the 
PREX-1 and -2 combined data is determined as 0.283$\pm$0.071 fm~\cite{Adhik2021PRL}. 
Within a specific class of relativistic energy density functionals,
Reed \textit{et al.} determines  $L=(106\pm 37)$ MeV~\cite{ReedB2021PRL}, implying a stiff symmetry energy at the typical densities of $\rho \sim 0.15$ fm$^{-3}$ in atomic nuclei. It is systematically larger than previous limits based on both theoretical approaches and experimental measurements.
From the linear correlation between $L$ and neutron skin thickness of $^{208}$Pb within a class of extended Skyrme density functionals, Yue \textit{et al.}~\cite{YueGT2021arXiv} gives $L=85.5\pm 22.2$ MeV by the constraints from the PREX experiment and the symmetry energy density slope $L(\rho)$ values at subsaturation cross density $\rho_c$=0.11/0.16 $\rho_0$. 
A reanalysis of the PREX data based on different types of nuclear energy density functions (EDF) gives $L=54\pm8$ MeV~\cite{Reinh2021PRL}. The discrepancy comes from the fact that the significant one-sigma uncertainty of PREX-2 value of the parity-violating asymmetry (APV) makes it difficult to use it to constraint current EDFs.

Mirror nuclei are the nuclei with the same atomic number $A$ but with the proton number $Z$ and neutron number $N$ interchanged.  
Assuming a perfect charge symmetry in atomic nuclei, the point-proton density distribution $\rho_p$ of one nucleus $(Z, N)$, would be the same as the point-neutron density distribution $\rho_n$ of the mirror nuclide ($N, Z$). 
As a step further, one expects that 
the proton distribution radius of one nucleus $(Z, N)$ would be the same as  the 
neutron distribution radius of the mirror nuclide ($N, Z$). 
Then the difference of proton radii between mirror nuclei, $R_\text{mirr}$, should naturally reflect the neutron skin thickness of ($Z, N$), $R_{\text{skin}}(Z,N)$, namely,  
\begin{equation}\label{2}
\begin{aligned}
R_{\text{skin}}(Z,N)&\equiv R_{n}(Z,N)-R_{p}(Z,N)\\
&\approx R_{p}(N,Z)-R_{p}(Z,N)\equiv R_\text{mirr}(Z,N).
\end{aligned}
\end{equation}
 
The linear correlation between the charge radius difference of mirror nuclei, $R_\text{mirr}$, and the slope parameter $L$ was first identified for the mirror pair $^{30}$S and $^{30}$Si by Wang \textit{et al.}~\cite{wangN2013PRC}. Brown~\cite{Brown2017PRL} examined by constructing 48 Skyrme functionals that the charge radius differences of mirror nuclei are proportional to $L$ at saturation density for a particular mirror pair, 
even in the presence of the Coulomb correction. 
The same findings were confirmed later in an approach based on the
relativistic mean-field model spanning a large range of the slope of the symmetry energy~ $L$~\cite{YangJ2018PRC,Gaida2020NPA}. 
Given the  $R_\text{mirr}$ – $L$ correlation at the nuclear saturation density, Brown~\cite{Brown2020PRR} deduced $L = 5$–$70$ MeV 
based on charge radii of the $^{36}$Ca -$^{36}$S and $^{38}$Ca -$^{38}$Ar mirror pairs. 
They found a different slope for the linearity by comparing the theoretical predictions based on Skyrme and Relativistic mean field functionals.  

Experimentally, the measurement of charge radii of atomic nuclei can be performed by electron scattering ~\cite{Abrah2012PRL,Jager1974ADNDT,sudaT2017PPNP}, $\mu$ atomic spectrum~\cite{engfe1974ADNDT,Frick1995ADNDT}, laser spectroscopy~\cite{LuZ-T2013RMP,campb2016PPNP}. However, due to factors such as the yield of unstable nuclei and the requirement of radioactive isotopic targets, all these methods have certain limitations when studying peculiar nuclei far away from the stability line. 
Recently, a new and effective method has been developed to extract charge radii of unstable nuclei~\cite{Yamag2010PRC,Yamag2011PRL}.
It relies on the precise charge-changing cross section measurements $\sigma_\text{cc}$ on a well known target in combination with Glauber-type model interpretation.
This method is suitable to extract the proton radii of radioactive isotopes, and 
has been used in the last years to determine the proton radii of several isotopic chains, e.g., $^{12-19}$C~\cite{Yamag2011PRL,Kanun2016PRL},  $^{7,9-12,14}$B~\cite{Terash2014PTEP}, $^{17-22}$N~\cite{Bagchi2019PLB}, $^{30}$Ne and $^{32,33}$Na~\cite{Ozawa2014PRC}.
  
The concrete purpose of this work is essentially twofold: One intention is to propose a potential new probe to study $L$ from the aspect of charge-changing cross section measurements of mirror nuclei. This method is easy to evaluate and 
has a similar precision to the neutron skin thickness $R_{\text{skin}}$ and the charge radius difference $R_\text{mirr}$ of mirror nuclei. 
The second purpose is that we will discuss how robust of this new probe regards to reaction energies and masses of mirror nuclei. 

\section{Charge-changing cross section difference of mirror nuclei}
\label{Section:ExpResults}

The charge-changing cross section ($\sigma_\text{cc}$) reflects the probability to remove one or more protons from the incident nucleus during a collision with a target nuclide. % Theoretically, it can be described by Glauber model when the collision energy is high.
Assuming that only the  protons in the projectile will contribute to the charge changing reaction, the experimental data at a few hundred MeV/nucleon or higher can be reasonably described by the zero-range optical-limit Glauber model~\cite{MengJ2002PLB,Bhagw2004PRC,Yamag2010PRC}.  
%by scaling with a weak energy dependent factor~~\cite{Yamag2011PRL,Yamag2010PRC,Ozawa2014PRC}. 
 \iffalse
 The idea was to model the nucleus in the simplest way, as uncorrelated nucleons sampled from measured density distributions. Two nuclei could be arranged with a random impact parameter $b$
 and projected onto the x-y plane. Then interaction probabilities could be applied by using the relative distance between two nucleon centroids as a stand-in for the measured nucleon-nucleon inelastic cross section.
 \fi
Glauber model is based on the individual nucleon-nucleon collisions in the overlap volume of the high energy colliding nuclei, and the nuclide is modeled as uncorrelated nucleons sampled from the density distributions.
In this framework, 
$\sigma_\text{cc}$ can be written as:
\begin{center}
\begin{equation}
\sigma_\text{cc}=\int{(1-T^{p}(\boldsymbol b))d\boldsymbol b} \;,
\end{equation}
\label{eq:cc}    
\end{center} 
where $\boldsymbol b$ and $T^{p}(\boldsymbol b)$ are the impact parameter and and the transmission function, respectively.

$T^{p}(\boldsymbol b)$ reads  
\begin{center}
\begin{equation}
\begin{split}
T^{p}(b)=& exp\Big\{-\sigma_{pp}\int{[\rho^{\text{P}}_{p}(\boldsymbol{b-s})\cdot\rho^{\text{T}}_{p}(\boldsymbol{s})}]d\boldsymbol{s}\\
& -\sigma_{pn}\int{[\rho^{\text{P}}_{p}\boldsymbol{(b-s)}\cdot\rho^{\text{T}}_{n}(\boldsymbol{s})}]d\boldsymbol{s}\Big\} \;.
\label{eq:Tp}
\end{split}
\end{equation}
\end{center} 

Here $\sigma_{pp}$ and $\sigma_{pn}$ represent the free proton-proton and proton-neutron reaction cross section, respectively. \textbf{\textit{s}} and \textbf{\textit{(b-s)}} are the distances from the centers of the target and projectile nuclei, respectively.
$\rho^{\text{T(P)}}_{p(n)}$ is the proton (neutron) density distribution in target (projecitle) nuclide integrated along the beam axis.
Reaction targets like carbon or hydrogen are routinely used in the cross section measurements and have well established density distributions. $\rho^{\text{P}}_{p}$ of incident nuclei and $\sigma_\text{cc}$ are thus correlated. In case of a proton target, the proton distribution density can be described by a Dirac density function $\delta$. Eq.~(\ref{eq:Tp}) can be simplified as: 
\begin{center}
\begin{equation}
\begin{split}
T^{p}(b)=& exp\Big\{-\sigma_{pp}\int{[\rho^{\text{P}}_{p}(\boldsymbol{b-s})\cdot\delta^{\text{T}}_{p}(\boldsymbol{s})}]d\boldsymbol{s}\Big\}\\
=& exp[-\sigma_{pp}\cdot\rho^{\text{P}}_{p}(\boldsymbol{b})] \;.
\end{split}
\end{equation}
\label{eq:Tp_H}
\end{center} 

Calculations of $\sigma_\text{cc}$ depend on the accuracy in the physical inputs, and moreover how well one can treat the collision process of projectile protons with target nucleons in terms of nucleon-nucleon interactions.
Effects induced by, e.g., the zero-range approximation and the role of neutrons in the projectile nuclei have been discussed~\cite{Auman2017PRL,Abdul2020NPA,Suzuk2016PRC,Tanaka2021CCCS}.
However, in the present work, the absolute cross sections are less significant.
Instead, we focus on the difference of the charge-changing cross section ($\Delta\sigma_\text{cc}$) of mirror nuclei on the same target at the same energy. 
%\textcolor{red}{Meanwhile, we compared the relative errors of $\sigma_\text{cc}$ and $\Delta\sigma_\text{cc}$ between the zero and finite range. For $\sigma_\text{cc}$, the uncertainties caused by different Glauber model interpretation are reduced 3.51-74.99\% (from the different density distributions)in $\Delta\sigma_\text{cc}$. We would like to mention that all the results shown in this work remain valid when using finite range.
In this case, all the physical inputs are identical except the proton density distributions in projectiles. The uncertainties induced by the common parts, e.g., the density distribution in target nucleon, would be largely cancelled out. To quantify this, we refer to an empirical formula of $\sigma_\text{cc}$~\cite{Chuk2000NPA},
\begin{center}
\begin{equation}
\begin{split}
\sigma_{cc}=\pi [R^{\text{mat}}_{\text{targ}}+R^{\text{prot}}_{\text{proj}}+a \frac{R^{\text{mat}}_{\text{targ}} R^{\text{prot}}_{\text{proj}}}{ R^{\text{mat}}_{\text{targ}}+R^{\text{prot}}_{\text{proj}}}-c]^2 \;,
\label{eq:Rp_RT}
\end{split}
\end{equation}
\end{center} 
where $R^{\text{mat}}_{\text{targ}}$ and $R^{\text{prot}}_{\text{proj}}$ are the matter radius of the target nuclide and the radius of the proton distribution of projectile nuclide, respectively. $a$ is related to the surface interaction term and $c$ reflects the effect of surface transparency. 
%Based on Eq.~(\ref{eq:Rp_RT}), and the contribution of higher-order terms of $R_{\text{mirr}}$ is ignored, 
Considering the charge symmetry in mirror nuclei, $\Delta\sigma_{cc}$ can be approximated as:
\begin{center}
\begin{equation}
\begin{split}
\Delta\sigma_{cc} \approx 2\pi  (1+af^2)[R^{\text{mat}}_{\text{targ}}+afR^{\text{prot}}_{\text{proj}}(Z,N)-c] R_{\text{mirr}} \;,
\label{eq:deltacc_RT}
\end{split}
\end{equation}
\end{center} 
where $f=R^{\text{mat}}_{\text{targ}}/[R^{\text{mat}}_{\text{targ}}+R^{\text{prot}}_{\text{proj}}(Z,N)]$. To the first order, $\Delta\sigma_\text{cc}$ is nearly linearly correlated to $R_{\text{mirr}}$ and thus the neutron skin thickness. Moreover, the uncertainty induced by the matter radius of the target nuclide, is much reduced considering the small $R_{\text{mirr}}$.

\section{RESULTS AND DISCUSSIONS}
\label{section3}

%%%%%%%%%%%%%%%%%%%%%%%%%%%%%%%%%%%%%%%%%%%%%%%%%%%%%%%%%%%%%%%%%%%%%%%%%%%%%%%%%%%%%%%%%%%%%%%%%%
The ground-state properties of mirror nuclei including the radii, the proton and neutron density distributions and the slope of the symmetry energy are computed by the Skyrme-Hartree-Fock theory (SHF) with 39 different Skyrme forces~\cite{Chaba1997NPA,Chaba1998NPA,Reinh1995NPA,Klupf2009PRC,Barte1982NPA,Beine1975NPA,Krivi1980NPA}, including 
SLy, SkI, SV and SIII-SVI series.
%{\color {red} the 39 interaction forces are grouped according to the parametersets, SLy~\cite{1998Erratum,1997A}, Skl~\cite{1995Nuclear}, SV~\cite{Kl2009Variations} and SIII-SVI, etc~\cite{1982Towards,M1975Nuclear,M2012Skyrme,1980Derivation}.}
We take $^{30}$Si-$^{30}$S pair to study the relation between $\sigma_\text{cc}$ and $L$. 
The proton skin thickness of $^{30}$S is about a factor of 2 thicker than that of the neutron skin thickness of $^{30}$Si in the mean-field calculations. This interesting phenomenon~\cite{YangJ2019PRC} is interpreted as the enhanced effect in the neutron deficient nuclei due to both the Coulomb repulsion and the symmetry energy, which together push protons out to the surface and thus create a larger proton skin.

\begin{figure}[h]
\includegraphics[width=0.4\textwidth]{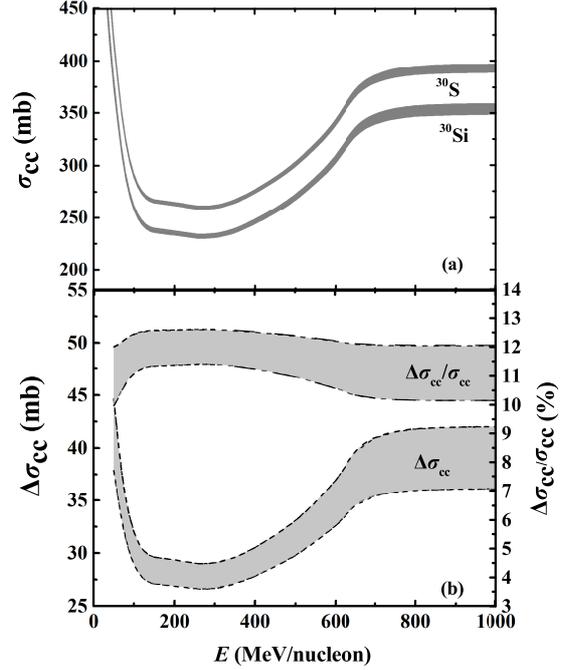}
\centering
\caption{(a) Charge-changing cross section for  $^{30}$Si and $^{30}$S on a proton target as functions of beam energy.
(b) Same as (a) but for the difference $\Delta\sigma_\text{cc}$ and the relative difference $\Delta\sigma_\text{cc}/\sigma_\text{cc}$. The band represents the variance when using the different proton density distributions in the SHF models.  }
\label{fig:cccs-E}
\end{figure}

We calculate the $\sigma_\text{cc}$ of $^{30}$Si and $^{30}$S on the $^{1}$H target using the zero-range optical limit Glauber model. 
A proton target is suggested to be a better choice to extract the nuclear size~\cite{Horiu2014PRC,Horiu2016PRC,Horiu2020PRC}. Other targets, like the commonly used carbon target, would complicate the present study of interest due to the additional contribution of neutrons in target nuclei. We would like to mention that all the results shown in this work remain valid when using a carbon target, although a slightly worse linearity with $L$ is found.

\begin{figure}[ht]
\includegraphics[width=0.35\textwidth]{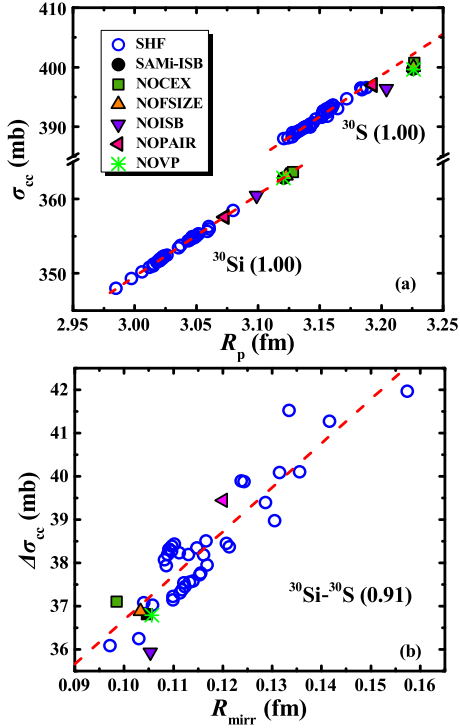}
\centering
\caption{(Color online) 
Correlation between $\sigma_\text{cc}$ and $R_{p}$ for $^{30}$Si and $^{30}$S (a), and that between the difference in cross sections ($\Delta \sigma_\text{cc}$) and in $rms$ proton radii ($R_\text{mirr}$) (b). The charge-changing cross sections are calculated at 900 MeV/nucleon on a proton target.
The dashed line is the linear fit to the calculated results by SHF. Numbers in parentheses represent the correlation coefficients. Refer to the text for details.} 
\label{fig:cccs-rp}
\end{figure} 

The charge-changing cross sections of mirror nuclei and their difference versus the incident energy are plotted in Fig.~\ref{fig:cccs-E}. The band represents the variation caused by choosing different proton density distributions in the SHF models. As expected, the energy evolution follows the same trend as $\sigma_{pp}$. The cross section of $^{30}$S is systematically larger than that of $^{30}$Si due to the enhanced collision of additional 2 protons in $^{30}$S. The cross section difference is minimal at an energy of around 250 MeV/nucleon, and keep almost constant at the energies of 700-1000 MeV/nucleon. The relative difference, $\Delta\sigma_\text{cc}/\sigma_\text{cc}$, on the other hand, reaches the maximum value of about 13\% at 250 MeV/nucleon, and has the maximum dispersion at 700-900 MeV/nucleon.  
Hereafter we take 900 MeV/nucleon as an illustration for further discussions.
At this energy, 
$\Delta\sigma_\text{cc}$ is more than 40\% larger than 250 MeV/nucleon, thus is 
easier experimentally. Furthermore, the maximum dispersion makes it best to distinguish the impact induced when using various proton density distributions. 

In Fig.~\ref{fig:cccs-rp}, we draw the scatter plots of $\sigma_\text{cc}$ versus root-mean-square ($rms$) proton distribution radii ($R_{p}$) for $^{30}$Si and $^{30}$S. 
In a geometric collision model, $\sigma_\text{cc}$ should be  approximately correlated to the square of $R_{p}$.
To the first order, $\sigma_\text{cc}$ versus $R_{p}$ follows well a straight line with a linear correlation coefficient of about 1.00. This results in the linear correlation between the cross section difference $\Delta \sigma_\text{cc}$ and the proton radius difference $R_\text{mirr}$ for the wide range of Skyrme parametrizations. 
The correlation is in a qualitative agreement with the linear fit to the microscopic calculations and Eq.~(\ref{eq:deltacc_RT}) as shown in Fig.~\ref{fig:cccs-rp}(b), namely, $\Delta\sigma_\text{cc}$=102.0(77)~$R_\text{mirr}$ + 26.5(9) (mb), 
 where $R_\text{mirr}$ is in the unit of fm. The corresponding linear correlation coefficient $r$ is 0.91.

\begin{figure*}[ht]
\includegraphics[width=0.8\textwidth]{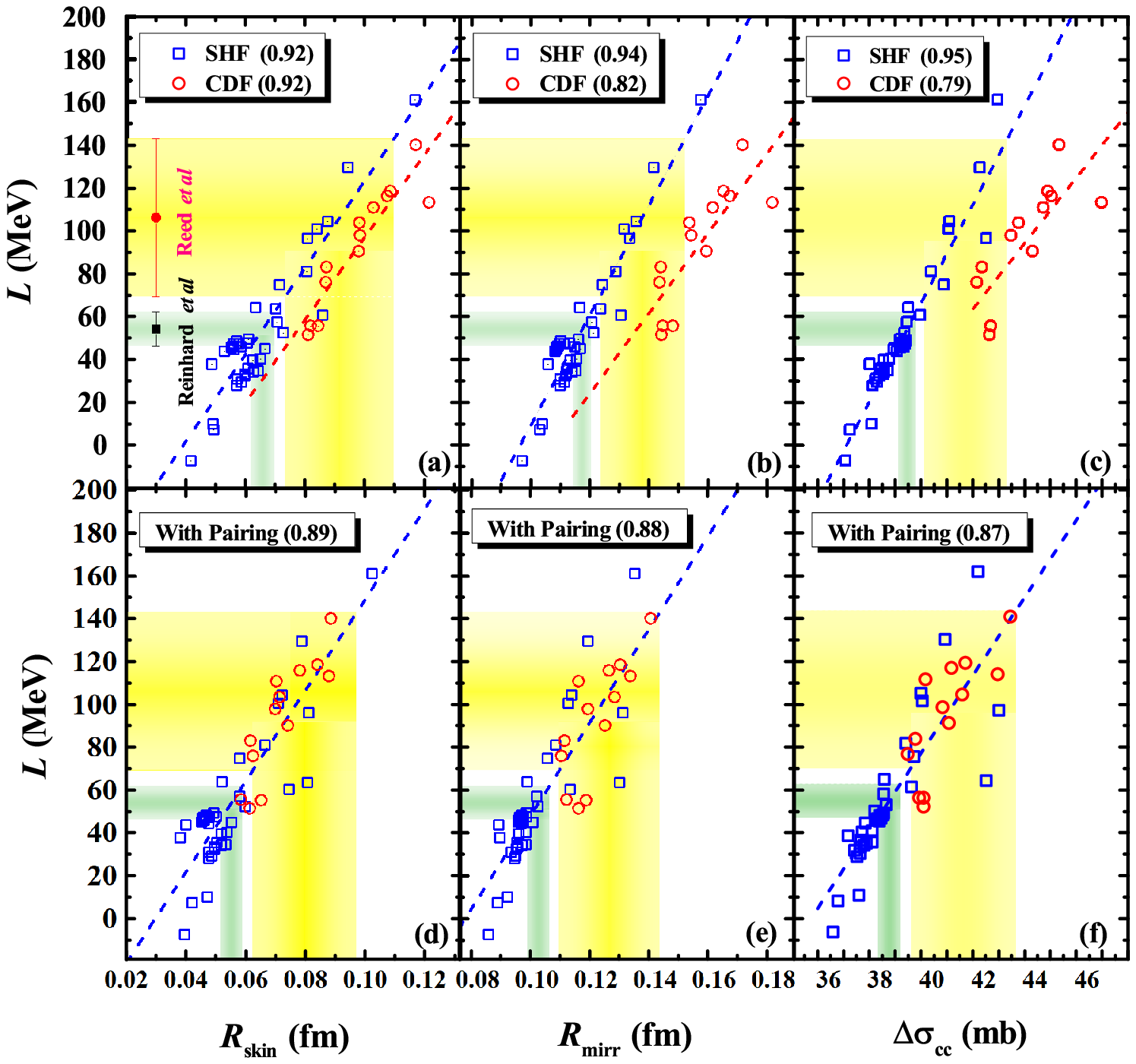}
\centering
\caption{ 
Slope of the symmetry energy at nuclear 
saturation density as a function of $R_{\text{skin}}$ (a), $R_\text{mirr}$ (b) and $\Delta\sigma_\text{cc}$ (c) for the $^{30}$Si-$^{30}$S pair. 
Panels (d)-(f) are the same as (a)-(c), but for including the pairing correction. The dashed lines are the relevant linear fits to SHF or/and CDF calculations.
 Numbers in parentheses represent the correlation coefficients. Indicated by the color coded areas are the two recent interpreted $L$ results from the combined PREX data. Refer to the text for details.}
\label{fig:deltacc-L}
\end{figure*}

Moreover, we performed the dedicated calculations using the SAMi-ISB forces~\cite{XRoca2018PRL}, an effective Skyrme-like interaction that
reconciles standard nuclear ground-state properties with the current understanding of the density behavior of the symmetry energy and the reproduction of the isobaric analogue state (IAS) energy of $^{208}$Pb as well as in Sn isotopes.  The NOCEX, NOFSIZE, NOISB, NOPAIR and NOVP refer to the calculations when neglecting Coulomb exchange, electromagnetic finite size effects, charge-symmetry-breaking (CSB) and charge-independence-breaking (CIB) effects from nuclear strong force, pairing and vacuum polarization correction to the Coulomb potential, respectively.
Nicely, the results obtained by switching off these corrections in strong and Coulomb forces, have little impact on the general linear trend from the wide range of Skyrme parametrizations.
 
%This assesses the robustness of the correlation between $\Delta\sigma_\text{cc}$ and $R_\text{mirr}$. With a good linear relationship between $R_\text{mirr}$ and $L$~\cite{wangN2013PRC}, establishing %the relationship between $\Delta\sigma_\text{cc}$ and $L$ is promising.}

In Fig.~\ref{fig:deltacc-L}, the density dependence of symmetry energy, $L$ are shown as a function of $R_{\text{skin}}$, $R_\text{mirr}$, and $\Delta\sigma_\text{cc}$ for $^{30}$Si-$^{30}$S pair, respectively. 
 Within the Skyrme functionals, the linear correlation coefficients of $L$ versus $\Delta\sigma_\text{cc}$ reaches 0.95. It is comparable to 0.92 for $L$ versus $R_{\text{skin}}$ and 0.94 for $L$ versus $R_\text{mirr}$.
Considering the fact that a direct measurement of its charge radius or the neutron skin thickness remain a challenge for isotopes like Si, $\Delta\sigma_\text{cc}$ can be used as a new probe to constraint $L$. It avoids the model uncertainty rooted in the radius determination from the Glauber-type model. 
The very different $L$ predictions from Ref.~\cite{Adhik2021PRL,ReedB2021PRL} and Ref.~\cite{Reinh2021PRL} are also indicated by the color-coded areas in this figure. 
To constrain $L$, a precision down to 1 mb is required. This is feasible experimentally because many systematic errors can be reduced in the difference for mirror nuclei.  
 
We extended the calculations by using the covariant (relativistic) density functionals (CDF). The widely used parameter sets, 
nonlinear parameter sets NL1, 2~\cite{Lalaz1997PRC,Sugah1994NPA}, density-dependent RMF (DDRMF) parameter sets DD-ME1, 2, PKDD ~\cite{Nikss2002PRC,Lalaz2005PRC,LongW2004PRC,Typel1999NPA}, and density-dependent relativistic Hartree-Fock (DDRHF) parameter sets PKO1, 2, 3~\cite{LongW2007PRC,LongW2008EPL} are shown in Fig.~\ref{fig:deltacc-L}.
The relativistic results generally scatter along a linear trend  and show a somewhat different slope from the SHF  results.  Taking $\Delta\sigma_\text{cc}=42.0 \pm 0.5 $ mb 
 as an example, the SHF and CDF  results would give $L=135 \pm 11$ and $79 \pm 8$ MeV, respectively. The SHF prediction is about 71\% larger than the CDF one.  
 
The different linearities in the relativistic and non-relativistic models in fact reflect the different structures in prediction of the mirror nuclei, as noticed for example in Ref.~\cite{Haruk2021PTEP}. We have performed a check on the effect of spin-orbit interactions similar to Ref.~\cite{Brown2020PRR} on the central bump of the density for $N$=16 and $Z$=16. Spin-orbit effects in this pair of $^{30}$Si-$^{30}$S has some impact, but it can not resolve the discrepancy between relativistic and non-relativistic models. For open shell nuclei, pairing correlations should be taken into account~\cite{Reinh2022PRC}. 
We have fitted for each functional the neutron and proton pairing gaps to the experimental values evaluated by the five-point mass-difference formulae along the $N$, $Z$=14 and $N$, $Z$=16 isotopic and isotonic chains. 
 This will allow to fix the pairing strength to a realistic value and avoid the problems when magic numbers appear in the calculations.
The new results with the inclusion of pairing correlations are shown accordingly in Fig.~\ref{fig:deltacc-L} (d)-(f). Now the SHF and CDF results are consistent in describ $L$ versus $R_\text{skin}$, $R_\text{mirr}$ and $\Delta\sigma_\text{cc}$. The resulting linear correlation coefficient is somehow in between the values of SHF and CDF without the pairing.
%As an example, $L=28.6(22)~\Delta\sigma_\text{cc}$ - 957 (78) (MeV), where $\Delta\sigma_\text{cc}$ is in the unit of mb. 
The linear coefficient of $L$ versus $\Delta\sigma_\text{cc}$ amounts to be 0.87, which is the same as those found between $L$ and $R_\text{skin}$ or $R_\text{mirr}$.
Moreover, we performed the calculations of $\Delta\sigma_\text{cc}$ in the energy range of 100 to 900 MeV/nucleon, and obtained a similar linearity with $L$.

\begin{figure}[H]
\includegraphics[width=0.45\textwidth]{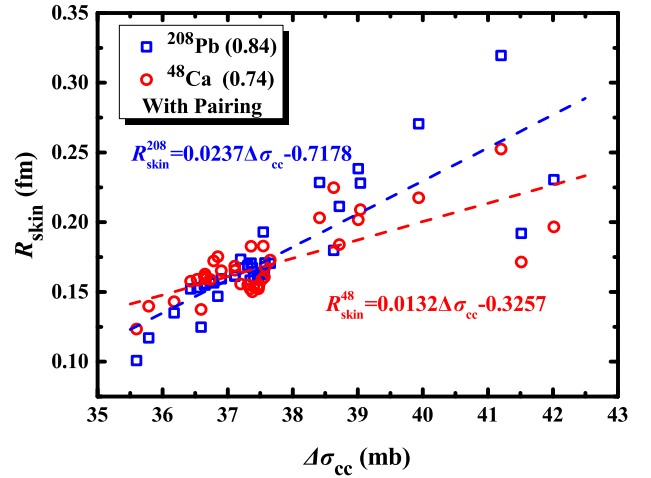}
\centering
\caption{ Relations between the neutron-skin thickness of $^{208}$Pb, $^{48}$Ca and the difference in charge-changing cross sections of $^{30}$Si-$^{30}$S at 900 MeV/nucleon. Numbers in parentheses represent the correlation coefficients. The linear fits are also shown. }
\label{fig:208Pb}
\end{figure}

One may question if the light mirror nuclei can be used to constrain EoS, since they are far from the infinite matter. The infinite matter refers to an infinite uniform system of nucleons, which is via the strong force without electromagnetic interactions. 
In addition, the surface corrections could be large in light nuclei. 
Answering this question has to rely on the microscopic models: can they describe the light, intermediate, and heavy nuclei self-consistently with a good accuracy? 
 The neutron skin thicknesses of $^{48}$Ca and $^{208}$Pb are well known to be linearly correlated with $L$ at the nuclear matter saturation density in e.g.,~\cite{Furns2002NPA,Chenl2005PRC,Cente2009PRL,Korte2013PRC,Reinh2016PRC}. In Fig.~\ref{fig:208Pb}, with the same SHF interaction
forces but including the pairing correlations, we examine the relations between $\Delta\sigma_\text{cc}$ of $^{30}$Si-$^{30}$S mirror nuclei and the neutron-skin thickness of $^{48}$Ca and $^{208}$Pb.   
We see a linear correlation for both $^{48}$Ca and
$^{208}$Pb, and a coefficient of  0.84 is reached for the latter.  
This indicates that even the light-mass mirror pair, although far away from the infinite matter, informs us about the density dependence of the symmetry energy around saturation. However, it should be noted including the pairing correlations reduces slightly the linearity with $L$.

Having established the existence of a strong correlation between $\Delta\sigma_\text{cc}$ and $L$, we discuss now the experimental feasibility. Due to the fact that systematic uncertainties associated to the large set of nuclear models employed will dominate with respect to pairing or deformation effects, we will neglect such effects in our qualitative discussion. Specifically, we calculate $\Delta\sigma_\text{cc}$ of 16 pairs of mirror nuclei from $^{14}$C-$^{14}$O to $^{58}$Ni-$^{58}$Zn.  
Both$\Delta\sigma_\text{cc}$ and $\Delta\sigma_\text{cc}/\sigma_\text{cc}$ decrease with the increasing proton number. The relative variance varies from about 29\% at $A=20$ to 10\% at $A=52$ for the $T_Z$=2 chain, and from about 22\% at $A=14$ to about 5\% at $A=58$ for the $Tz=1$ chain. The lighter the mirror nuclei are, the easier the experiment can be carried out with the present experimental technique.

In the current charge-changing cross section measurements, the typical uncertainty for charge radius extraction is a few mb and the relative uncertainty can be as precise as 1\% or even lower provided a good statistic. Therefore, it is experimentally reachable in the current experimental studies. We noted that a similar high precision in total neutron-removal cross sections has been proposed to study $L$ in Ref.~\cite{Auman2017PRL}. 
As for the isotopic chain, the slope of the $T_Z=2$ chain as a function of mass number $A$ is about a factor of 2 larger than that for the $T_Z=1$ chain. This is simply due to the enhanced effect from the proton-neutron asymmetry. 
A particularly interesting point would be to perform a systematic evaluation for various mirror nuclei, helping to verify the new $L$ probe.

\section{Summary}
\label{Section 4}

We studies $\sigma_\text{cc}$ by using the mean field model SHF and CDF together with a Glauber model analysis.
In this work, we show there is a good linear correlation between the difference of charge-changing cross sections of mirror nuclei on the same reaction target at the same energy and the density dependence of symmetry energy. In the cross section calculations, the proton density distributions are taken from both the SHF and CDF models. 
The pairing correlation is found to be crucial to get 
linear correlations consistently of $R_\text{skin}$, $ R_\text{mirr}$, and $\Delta\sigma_\text{cc}$ to $L$. 
Our results improve the theoretical analysis presented in previous investigations~\cite{Brown2017PRL,YangJ2018PRC,Brown2020PRR}.
Our results also support the recent study~\cite{Reinh2022PRC} that neglecting pairing would artificially increase the correlation between $ R_\text{mirr}$ and $L$.

The present work shows the possibility of using the charge-changing reactions of mirror nuclei to probe the density dependence of the symmetry energy. A high precision measurement of $\sigma_\text{cc}$ is already possible  with the current facility.
 Thus, dedicated experiments have been already proposed at HIRFL-CSR~\cite{SunBH2018SC} and HIMAC. The energy-, mass-, and isospin-dependent evaluation are key issues to constrain the symmetry energy. Direct comparisons with experimental observable are expected soon.

\section*{\uppercase{acknowledgments}}
\label{acknowledgments}
We would like to thank the reviewers for valuable comments. This work was supported in part by the National Natural Science Foundation of China (Nos. U1832211, 11961141004, 11922501 and 12075104). We thank Dr. Andrej Prochazka for helpful discussions on the Glauber model.

%==========================================================%
%==================BIBLIOGRAPHY============================%
%==========================================================%
\bibliographystyle{elsarticle-num}
%\bibliography{Ref}

\begin{thebibliography}{10}
\expandafter\ifx\csname url\endcsname\relax
  \def\url#1{\texttt{#1}}\fi
\expandafter\ifx\csname urlprefix\endcsname\relax\def\urlprefix{URL }\fi
\expandafter\ifx\csname href\endcsname\relax
  \def\href#1#2{#2} \def\path#1{#1}\fi

\bibitem{Stein2006Phys.Rep}
A.~W. Steiner, M.~Prakash, J.~M. Lattimer, P.~J. Ellis, Phys. Rep. 411~(25)
  (2005) 325.

\bibitem{LiBao2008Phys.Rep}
B.~A. Li, L.~W. Chen, C.~M. Ko, Phys. Rep. 464~(4-6) (2008) 113--281.

\bibitem{Latti2007Phys.Rep}
J.~M. Lattimer, M.~Prakash, Phys. Rep. 442~(1-6) (2007) 109--165.

\bibitem{ShenH2020ApJ}
H.~Shen, F.~Ji, J.~Hu, K.~Sumiyoshi, Astrophys. J 891~(2) (2020) 148.

\bibitem{XRoca2018PPNP}
X.~Roca-Maza, N.~Paar, Prog. Part. Nucl. Phys. 101 (2018) 96--176.

\bibitem{Burgi2021PPNP}
G.~Burgio, H.-J. Schulze, I.~Vidaña, J.-B. Wei, Prog. Part. Nucl. Phys. 120
  (2021) 103879.

\bibitem{Baldo2016PPNP}
M.~Baldo, G.~Burgio, Prog. Part. Nucl. Phys. 91 (2016) 203--258.

\bibitem{Latti2016PR}
J.~M. Lattimer, M.~Prakash, Phys. Rep. 621 (2016) 127--164.

\bibitem{Zhang2020FoP}
Y.-X. Zhang, N.~Wang, Q.-F. Li, L.~Ou, J.-L. Tian, M.~Liu, K.~Zhao, X.-Z. Wu,
  Z.-X. Li, Front. Phys. 15 (2020) 54301.

\bibitem{LiHan2020PRL}
H.~Li, H.~J. Xu, Y.~Zhou, X.~B. Wang, J.~Zhao, L.~W. Chen, F.~Q. Wang, Phys.
  Rev. Lett. 125 (2020) 222301.

\bibitem{XuHao2021PLB}
H.~jie Xu, H.~Li, X.~Wang, C.~Shen, F.~Wang, Phys. Lett. B 819 (2021) 136453.

\bibitem{WanNi2016PRC}
N.~Wan, C.~Xu, Z.~Ren, Phys. Rev. C 94~(4) (2016) 044322.

\bibitem{Zhou2019ApJ}
Y.~Zhou, L.-W. Chen, Astrophys. J 886~(1) (2019) 52.

\bibitem{Baiot2019PPNP}
L.~Baiotti, Prog. Part. Nucl. Phys. 109 (2019) 103714.

\bibitem{Willi1966Nucl.Phys}
W.~D. Myers, W.~J. Swiatecki, Nucl. Phys. 81~(1) (1966) 1--60.

\bibitem{Pomor2003PRC}
K.~Pomorski, J.~Dudek, Phys. Rev. C 67 (2003) 044316.

\bibitem{Horow2001PRL}
C.~J. Horowitz, J.~Piekarewicz, Phys. Rev. Lett. 86 (2001) 5647--5650.

\bibitem{Tsang2012PRC}
M.~B. Tsang, J.~R. Stone, F.~Camera, P.~Danielewicz, S.~Gandolfi, K.~Hebeler,
  C.~J. Horowitz, J.~Lee, W.~G. Lynch, Z.~Kohley, R.~Lemmon, P.~M\"oller,
  T.~Murakami, S.~Riordan, X.~Roca-Maza, F.~Sammarruca, A.~W. Steiner,
  I.~Vida\~na, S.~J. Yennello, Phys. Rev. C 86 (2012) 015803.

\bibitem{Horow2014JPG}
C.~J. Horowitz, E.~F. Brown, Y.~Kim, W.~G. Lynch, R.~Michaels, A.~Ono,
  J.~Piekarewicz, M.~B. Tsang, H.~H. Wolter, J. Phys. G: Nucl. Part. Phys.
  41~(9) (2014) 093001.

\bibitem{Typel2001PRC}
S.~Typel, B.~A. Brown, Phys. Rev. C 64 (2001) 027302.

\bibitem{Reinh2016PRC}
P.-G. Reinhard, W.~Nazarewicz, Phys. Rev. C 93 (2016) 051303.

\bibitem{Tamii2011PRL}
A.~Tamii, I.~Poltoratska, P.~von Neumann-Cosel, Y.~Fujita, T.~Adachi, C.~A.
  Bertulani, J.~Carter, M.~Dozono, H.~Fujita, K.~Fujita, K.~Hatanaka,
  D.~Ishikawa, M.~Itoh, T.~Kawabata, Y.~Kalmykov, A.~M. Krumbholz,
  E.~Litvinova, H.~Matsubara, K.~Nakanishi, R.~Neveling, H.~Okamura, H.~J. Ong,
  B.~\"Ozel-Tashenov, V.~Y. Ponomarev, A.~Richter, B.~Rubio, H.~Sakaguchi,
  Y.~Sakemi, Y.~Sasamoto, Y.~Shimbara, Y.~Shimizu, F.~D. Smit, T.~Suzuki,
  Y.~Tameshige, J.~Wambach, R.~Yamada, M.~Yosoi, J.~Zenihiro, Phys. Rev. Lett.
  107 (2011) 062502.

\bibitem{Latti2013ApJ}
J.~M. Lattimer, Y.~Lim, Astrophys. J 771~(1) (2013) 51.

\bibitem{stein2012PRL}
A.~W. Steiner, S.~Gandolfi, Phys. Rev. Lett. 108 (2012) 081102.

\bibitem{iida2003PLB}
K.~Iida, K.~Oyamatsu, B.~Abu-Ibrahim, Phys. Lett. B 576~(3) (2003) 273--280.

\bibitem{Krasznahorkay1999}
A.~Krasznahorkay, M.~Fujiwara, P.~van Aarle, H.~Akimune, I.~Daito, H.~Fujimura,
  Y.~Fujita, M.~N. Harakeh, T.~Inomata, J.~J\"anecke, S.~Nakayama, A.~Tamii,
  M.~Tanaka, H.~Toyokawa, W.~Uijen, M.~Yosoi, Phys. Rev. Lett. 82 (1999)
  3216--3219.

\bibitem{Krasznahorkay2013}
A.~Krasznahorkay, N.~Paar, D.~Vretenar, M.~N. Harakeh, Phys. Lett. B 720 (2013)
  428--432.

\bibitem{Cao2015}
L.-G. Cao, X.~Roca-Maza, G.~Col\`o, H.~Sagawa, Phys. Rev. C 92~(3) (2015)
  034308.

\bibitem{Roca-Maza2016}
X.~Roca-Maza, L.-G. Cao, G.~Col\`o, H.~Sagawa, Phys. Rev. C 94~(4) (2016)
  044313.

\bibitem{Chang2010PRC}
C.~Xu, B.-A. Li, L.-W. Chen, Phys. Rev. C 82 (2010) 054607.

\bibitem{WangY2020PLB}
Y.~Wang, Q.~Li, Y.~Leifels, A.~{Le Fèvre}, Phys. Lett. B 802 (2020) 135249.

\bibitem{Essic2021PRL}
R.~Essick, I.~Tews, P.~Landry, A.~Schwenk, Phys. Rev. Lett. 127 (2021) 192701.

\bibitem{LiBao2008PLB}
B.~A. Li, X.~Han, Phys. Lett. B 727 (2013) 276--281.

\bibitem{Oerte2017RevModPhys}
M.~Oertel, M.~Hempel, T.~Kl\"ahn, S.~Typel, Rev. Mod. Phys. 89 (2017) 015007.

\bibitem{Brown2000PRL}
B.~Alex~Brown, Phys. Rev. Lett. 85 (2000) 5296--5299.

\bibitem{Chenl2005PRC}
L.-W. Chen, C.~M. Ko, B.-A. Li, Phys. Rev. C 72 (2005) 064309.

\bibitem{Cente2009PRL}
M.~Centelles, X.~Roca-Maza, X.~Vi$\widetilde{n}$as, M.~Warda, Phys. Rev. Lett.
  102~(12) (2009) 122502.

\bibitem{RocaM2011PRL}
X.~Roca-Maza, M.~Centelles, X.~Vi$\widetilde{n}$as, M.~Warda, Phys. Rev. Lett.
  106 (2011) 252501.

\bibitem{Adhik2021PRL}
D.~Adhikari, H.~Albataineh, D.~Androic, et~al., Phys. Rev. Lett. 126 (2021)
  172502.

\bibitem{ReedB2021PRL}
B.~T. Reed, F.~Fattoyev, C.~Horowitz, J.~Piekarewicz, Phys. Rev. Lett. 126~(17)
  (2021) 172503.

\bibitem{YueGT2021arXiv}
T.-G. Yue, L.-W. Chen, Z.~Zhang, Y.~Zhou, Constraints on the symmetry energy
  from prex-ii in the multimessenger era (2021).

\bibitem{Reinh2021PRL}
P.-G. Reinhard, X.~Roca-Maza, W.~Nazarewicz, Phys. Rev. Lett. 127 (2021)
  232501.

\bibitem{wangN2013PRC}
N.~Wang, T.~Li, Phys. Rev. C 88~(1) (2013) 83--93.

\bibitem{Brown2017PRL}
B.~A. Brown, Phys. Rev. Lett. 119~(12) (2017) 122502.

\bibitem{YangJ2018PRC}
J.~Yang, J.~Piekarewicz, Phys. Rev. C 97~(1) (2018) 014314.

\bibitem{Gaida2020NPA}
M.~Gaidarov, I.~Moumene, A.~Antonov, D.~Kadrev, P.~Sarriguren, E.~{Moya de
  Guerra}, Nucl. Phys. A 1004 (2020) 122061.

\bibitem{Brown2020PRR}
B.~A. Brown, K.~Minamisono, J.~Piekarewicz, H.~Hergert, D.~Garand, A.~Klose,
  K.~K\"onig, J.~D. Lantis, Y.~Liu, B.~Maa\ss{}, A.~J. Miller,
  W.~N\"ortersh\"auser, S.~V. Pineda, R.~C. Powel, D.~M. Rossi, F.~Sommer,
  C.~Sumithrarachchi, A.~Teigelh\"ofer, J.~Watkins, R.~Wirth, Phys. Rev.
  Research 2 (2020) 022035.

\bibitem{Abrah2012PRL}
S.~Abrahamyan, A.~Acha, A.~Afanasev, et~al., Phys. Rev. Lett. 109 (2012)
  192501.

\bibitem{Jager1974ADNDT}
C.~{De Jager}, H.~{De Vries}, C.~{De Vries}, At. Data Nucl. Data Tables 14~(5)
  (1974) 479--508.

\bibitem{sudaT2017PPNP}
T.~Suda, H.~Simon, Prog. Part. Nucl. Phys. 96 (2017) 1--31.

\bibitem{engfe1974ADNDT}
R.~Engfer, H.~Schneuwly, J.~Vuilleumier, H.~Walter, A.~Zehnder, At. Data Nucl.
  Data Tables 14~(5) (1974) 509--597.

\bibitem{Frick1995ADNDT}
G.~Fricke, C.~Bernhardt, K.~Heilig, L.~Schaller, L.~Schellenberg, E.~Shera,
  C.~Dejager, At. Data Nucl. Data Tables 60~(2) (1995) 177--285.

\bibitem{LuZ-T2013RMP}
Z.-T. Lu, P.~Mueller, G.~W.~F. Drake, W.~N\"ortersh\"auser, S.~C. Pieper, Z.-C.
  Yan, Rev. Mod. Phys. 85 (2013) 1383--1400.

\bibitem{campb2016PPNP}
P.~Campbell, I.~Moore, M.~Pearson, Prog. Part. Nucl. Phys. 86 (2016) 127--180.

\bibitem{Yamag2010PRC}
T.~Yamaguchi, M.~Fukuda, S.~Fukuda, G.~W. Fan, I.~Hachiuma, M.~Kanazawa,
  A.~Kitagawa, T.~Kuboki, M.~Lantz, M.~Mihara, M.~Nagashima, K.~Namihira,
  D.~Nishimura, Y.~Okuma, T.~Ohtsubo, S.~Sato, T.~Suzuki, M.~Takechi, W.~Xu,
  Phys. Rev. C 82 (2010) 014609.

\bibitem{Yamag2011PRL}
T.~Yamaguchi, I.~Hachiuma, A.~Kitagawa, K.~Namihira, S.~Sato, T.~Suzuki,
  I.~Tanihata, M.~Fukuda, Phys. Rev. Lett. 107 (2011) 032502.

\bibitem{Kanun2016PRL}
R.~Kanungo, W.~Horiuchi, G.~Hagen, G.~R. Jansen, P.~Navratil, F.~Ameil,
  J.~Atkinson, Y.~Ayyad, D.~Cortina-Gil, I.~Dillmann, A.~Estrad\'e,
  A.~Evdokimov, F.~Farinon, H.~Geissel, G.~Guastalla, R.~Janik, M.~Kimura,
  R.~Kn\"obel, J.~Kurcewicz, Y.~A. Litvinov, M.~Marta, M.~Mostazo, I.~Mukha,
  C.~Nociforo, H.~J. Ong, S.~Pietri, A.~Prochazka, C.~Scheidenberger, B.~Sitar,
  P.~Strmen, Y.~Suzuki, M.~Takechi, J.~Tanaka, I.~Tanihata, S.~Terashima,
  J.~Vargas, H.~Weick, J.~S. Winfield, Phys. Rev. Lett. 117 (2016) 102501.

\bibitem{Terash2014PTEP}
S.~Terashima, I.~Tanihata, R.~Kanungo, A.~Estradé, W.~Horiuchi, F.~Ameil,
  J.~Atkinson, Y.~Ayyad, D.~Cortina-Gil, I.~Dillmann, A.~Evdokimov, F.~Farinon,
  H.~Geissel, G.~Guastalla, R.~Janik, M.~Kimura, R.~Knoebel, J.~Kurcewicz,
  Y.~A. Litvinov, M.~Marta, M.~Mostazo, I.~Mukha, T.~Neff, C.~Nociforo, H.~J.
  Ong, S.~Pietri, A.~Prochazka, C.~Scheidenberger, B.~Sitar, Y.~Suzuki,
  M.~Takechi, J.~Tanaka, J.~Vargas, J.~S. Winfield, H.~Weick, Prog. Theor. Exp.
  Phys. 2014 (2014) 101D02.

\bibitem{Bagchi2019PLB}
S.~{Bagchi}, R.~{Kanungo}, W.~{Horiuchi}, G.~{Hagen}, T.~D. {Morris}, S.~R.
  {Stroberg}, T.~{Suzuki}, F.~{Ameil}, J.~{Atkinson}, Y.~{Ayyad},
  D.~{Cortina-Gil}, I.~{Dillmann}, A.~{Estrad{\'e}}, A.~{Evdokimov},
  F.~{Farinon}, H.~{Geissel}, G.~{Guastalla}, R.~{Janik}, S.~{Kaur},
  R.~{Kn{\"o}bel}, J.~{Kurcewicz}, Y.~A. {Litvinov}, M.~{Marta}, M.~{Mostazo},
  I.~{Mukha}, C.~{Nociforo}, H.~J. {Ong}, S.~{Pietri}, A.~{Prochazka},
  C.~{Scheidenberger}, B.~{Sitar}, P.~{Strmen}, M.~{Takechi}, J.~{Tanaka},
  Y.~{Tanaka}, I.~{Tanihata}, S.~{Terashima}, J.~{Vargas}, H.~{Weick}, J.~S.
  {Winfield}, Phys. Lett. B 790 (2019) 251--256.

\bibitem{Ozawa2014PRC}
A.~Ozawa, T.~Moriguchi, T.~Ohtsubo, N.~Aoi, D.~Q. Fang, N.~Fukuda, M.~Fukuda,
  H.~Geissel, I.~Hachiuma, N.~Inabe, Y.~Ishibashi, S.~Ishimoto, Y.~Ito,
  T.~Izumikawa, D.~Kameda, T.~Kubo, T.~Kuboki, K.~Kusaka, M.~Lantz, Y.~G. Ma,
  M.~Mihara, Y.~Miyashita, S.~Momota, D.~Nagae, K.~Namihira, D.~Nishimura,
  H.~Ooishi, Y.~Ohkuma, T.~Ohnishi, M.~Ohtake, K.~Ogawa, Y.~Shimbara, T.~Suda,
  T.~Sumikama, H.~Suzuki, S.~Suzuki, T.~Suzuki, M.~Takechi, H.~Takeda,
  K.~Tanaka, R.~Watanabe, M.~Winkler, T.~Yamaguchi, Y.~Yanagisawa, Y.~Yasuda,
  K.~Yoshinaga, A.~Yoshida, K.~Yoshida, Phys. Rev. C 89 (2014) 044602.

\bibitem{MengJ2002PLB}
J.~Meng, S.-G. Zhou, I.~Tanihata, Phys. Lett. B 532~(3-4) (2002) 209--214.

\bibitem{Bhagw2004PRC}
A.~Bhagwat, Y.~K. Gambhir, Phys. Rev. C 69~(1) (2004) 014315.

\bibitem{Auman2017PRL}
T.~Aumann, C.~A. Bertulani, F.~Schindler, S.~Typel, Phys. Rev. Lett. 119 (2017)
  262501.

\bibitem{Abdul2020NPA}
I.~Abdul-Magead, B.~Abu-Ibrahim, Nucl. Phys. A 1000 (2020) 121804.

\bibitem{Suzuk2016PRC}
Y.~Suzuki, W.~Horiuchi, S.~Terashima, R.~Kanungo, F.~Ameil, J.~Atkinson,
  Y.~Ayyad, D.~Cortina-Gil, I.~Dillmann, A.~Estrad\'e, A.~Evdokimov,
  F.~Farinon, H.~Geissel, G.~Guastalla, R.~Janik, R.~Knoebel, J.~Kurcewicz,
  Y.~A. Litvinov, M.~Marta, M.~Mostazo, I.~Mukha, C.~Nociforo, H.~J. Ong,
  S.~Pietri, A.~Prochazka, C.~Scheidenberger, B.~Sitar, P.~Strmen, M.~Takechi,
  J.~Tanaka, I.~Tanihata, J.~Vargas, H.~Weick, J.~S. Winfield, Phys. Rev. C 94
  (2016) 011602.

\bibitem{Tanaka2021CCCS}
M.~Tanaka, M.~Takechi, A.~Homma, A.~Prochazka, M.~Fukuda, D.~Nishimura,
  T.~Suzuki, T.~Moriguchi, D.~S. Ahn, A.~Aimaganbetov, M.~Amano, H.~Arakawa,
  S.~Bagchi, K.~H. Behr, N.~Burtebayev, K.~Chikaato, H.~Du, T.~Fujii,
  N.~Fukuda, H.~Geissel, T.~Hori, S.~Hoshino, R.~Igosawa, A.~Ikeda, N.~Inabe,
  K.~Inomata, K.~Itahashi, T.~Izumikawa, D.~Kamioka, N.~Kanda, I.~Kato,
  I.~Kenzhina, Z.~Korkulu, Y.~Kuk, K.~Kusaka, K.~Matsuta, M.~Mihara, E.~Miyata,
  D.~Nagae, S.~Nakamura, M.~Nassurlla, K.~Nishimuro, K.~Nishizuka, K.~Ohnishi,
  M.~Ohtake, T.~Ohtsubo, S.~Omika, H.~J. Ong, A.~Ozawa, H.~Sakurai,
  C.~Scheidenberger, Y.~Shimizu, T.~Sugihara, T.~Sumikama, H.~Suzuki,
  S.~Suzuki, H.~Takeda, Y.~Tanaka, Y.~K. Tanaka, I.~Tanihata, T.~Wada,
  K.~Wakayama, S.~Yagi, T.~Yamaguchi, R.~Yanagihara, Y.~Yanagisawa, K.~Yoshida,
  T.~K. Zholdybayev, Charge-changing cross sections for $^{42\textrm{--}51}$ca
  and effect of charged-particle evaporation induced by neutron removal
  reaction (2021).

\bibitem{Chuk2000NPA}
L.~Chulkov, O.~Bochkarev, D.~Cortina-Gil, H.~Geissel, M.~Hellström, M.~Ivanov,
  R.~Janik, K.~Kimura, T.~Kobayashi, A.~Korsheninnikov, G.~Münzenberg,
  F.~Nickel, A.~Ogloblin, A.~Ozawa, M.~Pfützner, V.~Pribora, M.~Rozhkov,
  H.~Simon, B.~Sitár, P.~Strmen, K.~Sümmerer, T.~Suzuki, I.~Tanihata,
  M.~Winkler, K.~Yoshida, Nucl. Phys. A 674~(3) (2000) 330--342.

\bibitem{Chaba1997NPA}
E.~Chabanat, P.~Bonche, P.~Haensel, J.~Meyer, R.~Schaeffer, Nucl. Phys. A
  635~(1-2) (1997) 231--256.

\bibitem{Chaba1998NPA}
E.~Chabanat, P.~Bonche, P.~Haensel, J.~Meyer, R.~Schaeffer, Nucl. Phys. A
  643~(4) (1998) 441--441.

\bibitem{Reinh1995NPA}
P.-G. Reinhard, H.~Flocard, Nucl. Phys. A 584~(3) (1995) 467--488.

\bibitem{Klupf2009PRC}
P.~Kl\"upfel, P.-G. Reinhard, T.~J. B\"urvenich, J.~A. Maruhn, Phys. Rev. C 79
  (2009) 034310.

\bibitem{Barte1982NPA}
J.~Bartel, P.~Quentin, M.~Brack, C.~Guet, H.-B. Håkansson, Nucl. Phys. A
  386~(1) (1982) 79--100.

\bibitem{Beine1975NPA}
M.~Beiner, H.~Flocard, N.~{Van Giai}, P.~Quentin, Nucl. Phys. A 238~(1) (1975)
  29--69.

\bibitem{Krivi1980NPA}
H.~Krivine, J.~Treiner, O.~Bohigas, Nucl. Phys. A 336~(2) (1980) 155--184.

\bibitem{YangJ2019PRC}
J.~Yang, J.~A. Hernandez, J.~Piekarewicz, Phys. Rev. C 100 (2019) 054301.

\bibitem{Horiu2014PRC}
W.~Horiuchi, Y.~Suzuki, T.~Inakura, Phys. Rev. C 89 (2014) 011601.

\bibitem{Horiu2016PRC}
W.~Horiuchi, S.~Hatakeyama, S.~Ebata, Y.~Suzuki, Phys. Rev. C 93 (2016) 044611.

\bibitem{Horiu2020PRC}
W.~Horiuchi, Y.~Suzuki, T.~Uesaka, M.~Miwa, Phys. Rev. C 102 (2020) 054601.

\bibitem{XRoca2018PRL}
X.~Roca-Maza, G.~Col\`o, H.~Sagawa, Phys. Rev. Lett. 120 (2018) 202501.

\bibitem{Lalaz1997PRC}
G.~A. Lalazissis, J.~K\"onig, P.~Ring, Phys. Rev. C 55 (1997) 540--543.

\bibitem{Sugah1994NPA}
Y.~Sugahara, H.~Toki, Nucl. Phys. A 579~(3) (1994) 557--572.

\bibitem{Nikss2002PRC}
T.~Nik\ifmmode \check{s}\else \v{s}\fi{}i\ifmmode~\acute{c}\else \'{c}\fi{},
  D.~Vretenar, P.~Finelli, P.~Ring, Phys. Rev. C 66 (2002) 024306.

\bibitem{Lalaz2005PRC}
G.~A. Lalazissis, T.~Nik\ifmmode \check{s}\else
  \v{s}\fi{}i\ifmmode~\acute{c}\else \'{c}\fi{}, D.~Vretenar, P.~Ring, Phys.
  Rev. C 71 (2005) 024312.

\bibitem{LongW2004PRC}
W.~H. Long, J.~Meng, N.~V. Giai, S.-G. Zhou, Phys. Rev. C 69 (2004) 034319.

\bibitem{Typel1999NPA}
S.~Typel, H.~Wolter, Nucl. Phys. A 656~(3) (1999) 331--364.

\bibitem{LongW2007PRC}
W.~H. Long, H.~Sagawa, N.~V. Giai, J.~Meng, Phys. Rev. C 76 (2007) 034314.

\bibitem{LongW2008EPL}
W.~H. Long, H.~Sagawa, J.~Meng, N.~V. Giai, {EPL} (Europhysics Letters) 82~(1)
  (2008) 12001.

\bibitem{Haruk2021PTEP}
H.~Kurasawa, T.~Suda, T.~Suzuki, Prog. Theor. Exp. Phys. 2021~(1) (2020)
  013D02.

\bibitem{Reinh2022PRC}
P.-G. Reinhard, W.~Nazarewicz, Phys. Rev. C 105 (2022) L021301.

\bibitem{Furns2002NPA}
R.~Furnstahl, Nucl. Phys. A 706~(1) (2002) 85--110.

\bibitem{Korte2013PRC}
M.~Kortelainen, J.~Erler, W.~Nazarewicz, N.~Birge, Y.~Gao, E.~Olsen, Phys. Rev.
  C 88 (2013) 031305.

\bibitem{SunBH2018SC}
B.-H. Sun, J.-W. Zhao, X.-H. Zhang, L.-N. Sheng, Z.-Y. Sun, I.~Tanihata,
  S.~Terashima, Y.~Zheng, L.-H. Zhu, L.-M. Duan, L.-C. He, R.-J. Hu, G.-S. Li,
  W.-J. Lin, W.-P. Lin, C.-Y. Liu, Z.~Liu, C.-G. Lu, X.-W. Ma, L.-J. Mao,
  Y.~Tian, F.~Wang, M.~Wang, S.-T. Wang, J.-W. Xia, X.-D. Xu, H.-S. Xu, Z.-G.
  Xu, J.-C. Yang, D.-Y. Yin, Y.-J. Yuan, W.-L. Zhan, Y.-H. Zhang, X.-H. Zhou,
  Sci. Bull. 63~(2) (2018) 78--80.

\end{thebibliography}

\end{document}